\documentclass[12pt]{article}

\begin{document}
\begin{center}
{\bf Higher Derivative Fermionic Field Equation in the First Order Formalism }\\
\vspace{5mm}
 S. I. Kruglov \\
\vspace{5mm}
\textit{University of Toronto at Scarborough,\\ Physical and Environmental Sciences Department, \\
1265 Military Trail, Toronto, Ontario, Canada M1C 1A4}
\end{center}

\begin{abstract}
The generalized Dirac equation of the third order, describing
particles with spin 1/2 and three mass states, is analyzed. We
obtain the first order generalized Dirac equation in the
24-dimensional matrix form. The mass and spin projection operators
are found which extract solutions of the wave equation corresponding
to pure spin states of particles. The density of the electromagnetic
current is obtained, and minimal and non-minimal (anomalous)
electromagnetic interactions of fermions are considered by
introducing three phenomenological parameters. The Hamiltonian form
of the first order equation has been obtained.

PACS: 03.65Pm; 11.10Ef; 12.10Kt
\end{abstract}

\section{Introduction}

The standard model (SM) of electroweak interactions does not
explain the mass spectra of leptons and quarks, and contains many
free parameters. One of the most fundamental problems is the mass
generation of fermions. Therefore, it is important to investigate
the unified description of quarks and leptons and understand the
origin of fermion masses. The mass generation mechanism is beyond
the SM and is needed for the deeper understanding of lepton-quark
families.

We consider here the toy model based on the generalized Dirac
equation of the third order (see \cite{Barut}\footnote{In the work
\cite{Barut}, authors investigated the case of two nonzero masses
and one zero mass for the unified description of $e$,
$\mu$-leptons and neutrino.}) possessing flavor $S_3$ symmetry,
for particles with three nonzero mass states. One can speculate
that this equation may be applied for unified description of
lepton-quark generations (see \cite{Barut1}, \cite{Wilson},
\cite{Dvoeglazov}, \cite{Kruglov} for the case of two mass
states). According to the Barut approach \cite{Barut1}, such
equations may be treated as effective equations for partly
``dressed" fermions, and such schemes represent the
non-perturbative approach to quantum electrodynamics.

It should be noted that higher derivative (HD) field theories have
been investigated already. Some examples include generalized
electrodynamics \cite{Podolski}, renormalizable gravity theory in
four dimensions \cite{Stelle} and others. Therefore, it is motivated
to investigate HD field theories in four dimensions.

In this paper, we represent the third order (in derivatives)
equation in the form of the first order generalized Dirac equation,
obtain solutions in the form of projection operators and consider
minimal and non-minimal electromagnetic interactions.

The paper is organized as follows. In Sec. 2, we define the
generalized Dirac equation of the third order for fermions, obtain
projection operators extracting solutions of the HD equation, and
analyze a propagator of fields. The first order generalized
24-component Dirac equation is derived in Sec. 3. It is proven that
the mass spectrum of the 24-component matrix equation obtained and
the third order 4-component equation is the same. We obtain mass and
spin projection operators in Sec. 4. These projection operators
extract pure spin states of particles. Sec. 5 is devoted to
introducing minimal and non-minimal electromagnetic interactions of
fermions. In this section, we find also the electromagnetic current
density. We postulate the matrix equation with three
phenomenological parameters characterizing anomalous electromagnetic
fermion interactions and derive the quantum-mechanical Hamiltonian.
A conclusion is made in Sec. 6. In appendixes, we obtain useful
relations of $24\times 24$-matrices of the wave equation and spin
matrices.

The system of units $\hbar =c=1$ is chosen and euclidian metric is
used (see \cite{Ahieser}).

\section{ Field Equation of Third Order for Fermions }

We postulate the third order (in derivatives) field equation
describing fermions:
\begin{equation}
\left(\gamma_\mu\partial_\mu +m_1\right)
\left(\gamma_\nu\partial_\nu +m_2\right)
\left(\gamma_\alpha\partial_\alpha +m_3\right)\psi(x)=0 ,
\label{1}
\end{equation}
where $\partial_\nu =\partial/\partial x_\nu =(\partial/\partial
x_m,\partial/\partial (it))$, $\psi (x)$ is a bispinor; $m_1$,
$m_2$, $m_3$ are masses of fermions. As usual, repeated indices
imply a summation. The Dirac Hermitian matrices $\gamma_\mu $ obey
the commutation relations $\gamma_\mu \gamma_\nu +\gamma_\nu
\gamma_\mu =2\delta_{\mu\nu}$. It is obvious, that there are three
independent solutions of Eq. (1) corresponding to masses $m_i$
($i=1, 2, 3$). The bispinor $\overline{\psi
}(x)=\psi^{+}(x)\gamma_4$ ($\psi^{+}$ is the Hermitian-conjugated
bispinor) satisfies the equation
\begin{equation}
\overline{\psi }(x)\left(\gamma_\mu\overleftarrow{\partial}_\mu
-m_1\right) \left(\gamma_\nu\overleftarrow{\partial}_\nu
-m_2\right) \left(\gamma_\alpha\overleftarrow{\partial}_\alpha
-m_3\right) =0 , \label{2}
\end{equation}
where the derivatives $\overleftarrow{\partial} _\mu$ act on the
left-standing function. The discrete symmetries of the wave
function ${\psi }(x)$ directly follow from the Dirac theory.

Eq. (1) for positive energy, in momentum space, is given by
\begin{equation}
\left(i\widehat{p}+m_1\right)\left(i\widehat{p}+m_2\right)
\left(i\widehat{p}+m_3\right) \psi(p)=0  ,
 \label{3}
\end{equation}
where $\widehat{p}=\gamma_\mu p_\mu$, $p_\mu=(\textbf{p},ip_0)$. For
antiparticles one should make the replacement $p\rightarrow -p$.
From Eq. (3), one obtains the dispersion equation as follows:
\begin{equation}
\left(p^2+m_1^2\right)\left(p^2+m_2^2\right) \left(p^2+m_3^2\right)
=0  ,
 \label{4}
\end{equation}

The projection operator extracting solutions to Eq. (3),
corresponding to positive energy, is given by
\begin{equation}
\Lambda_+= \frac{(- i\widehat{p}+m_1)(- i\widehat{p}+m_2)(-
i\widehat{p}+m_3)}{2[m_1m_2m_3-p^2(m_1+m_2+m_2)]} .\label{5}
\end{equation}
It is easy to verify that the projection operator (5) obeys the
necessary equation $\Lambda_{+}^2= \Lambda_{+}$. It should be noted
that the projection operator $\Lambda_+$ is the solution to Eq. (3)
for any momentum $p$ obeying the dispersion relation (4), i.e. the
momentum $p$ is not specified here. The projection operator
$\Lambda_{+}$ corresponds to positive energy, and at the replacement
$p\rightarrow -p$, to negative energy of particles.

In the momentum space the propagator of the HD field is
\[
\Delta(p)=\frac{1}{\left(i\widehat{p}+m_1\right)\left(i\widehat{p}+m_2\right)
\left(i\widehat{p}+m_3\right)}
\]
\vspace{-7mm}
\begin{equation} \label{6}
\end{equation}
\vspace{-7mm}
\[
=\frac{(- i\widehat{p}+m_1)(- i\widehat{p}+m_2)(-
i\widehat{p}+m_3)}{\left(p^2+m_1^2\right)\left(p^2+m_2^2\right)
\left(p^2+m_3^2\right)} .
\]
The propagator (6) can be represented as a sum of Dirac propagators
\begin{equation}
\Delta(p)=\frac{A(- i\widehat{p}+m_1)}{\left(p^2+m_1^2\right)}
-\frac{B(- i\widehat{p}+m_2)}{\left(p^2+m_2^2\right)}+\frac{C(-
i\widehat{p}+m_3)}{\left(p^2+m_3^2\right)} , \label{7}
\end{equation}
where the coefficients $A$, $B$, and $C$ are given by
\[
A = \frac{1}{(m_2 - m_1)( m_3 - m_1)},~~~~B=\frac{1}{(m_2 - m_1)(
m_3 - m_2)} ,
\]
\vspace{-7mm}
\begin{equation} \label{8}
\end{equation}
\vspace{-7mm}
\[
 C =\frac{1}{(m_3 - m_2)( m_3 - m_1)}.
\]
Eq. (7) is valid only in the case $m_1\neq m_2\neq m_3$, i.e. masses
of the field are not degenerated. In the case $m_3>m_2>m_1$, the
coefficients $A$, $B$, and $C$ are positive, and the second
propagator in Eq. (7), corresponding to the mass $m_2$, has the
``wrong" sign $(-)$. This means that the HD field possesses two
physical states with masses $m_1$, $m_3$, and one Weyl ghost with
the mass $m_2$. The ghost gives the negative contribution to the
energy \cite{Wilson}, \cite{Villasenor}, and, as a result, the
Hamiltonian is not bounded from below. Therefore, to formulate the
quantum field theory one needs to introduce the indefinite metrics
\cite{Wilson}. As the propagator (7) is a sum of Dirac propagators,
the renormalization of the theory can be performed with the help of
the standard procedure of QED.

\section{24-Component First Order Relativistic Wave Equation}

For a convenience, we rewrite Eq. (1) in the form
\begin{equation}
\left(\gamma_\mu\partial_\mu \partial_\nu^2 +a\gamma_\mu\partial_\mu
+ b\partial_\nu^2 + c \right)\psi(x)=0 , \label{9}
\end{equation}
where $\partial_\nu^2=\partial_m^2-\partial_0^2$ $(m=1,2,3$,
$\partial_0=\partial/\partial t)$. The  coefficients
\begin{equation}
a = m_1 m_2 + m_1 m_3 + m_2 m_3 ,~~ b = m_1 + m_2 + m_3 ,~~ c = m_1
m_2 m_3 \label{10}
\end{equation}
are invariant under the permutation symmetry $S_3$.

Now we reformulate the third order Eq. (9) in the form of the first
order relativistic wave equation. It is useful for Lagrangian
formulation of the theory, obtaining the conserved currents, and
different calculations. Let us introduce the 24-dimensional function
\begin{equation}
\Psi (x)=\left\{ \Psi _A(x)\right\} =\left(
\begin{array}{cc}
\psi (x)\\
\widetilde{\psi} (x)\\
\psi _\mu (x)
\end{array}
\right) , \label{11}
\end{equation}
where $A=0, \widetilde{0}, \mu $, and
\[
\Psi_0 (x)=\psi (x) ,~~~~\Psi_{\widetilde{0}} (x)= \widetilde{\psi}
(x) \equiv \frac{1}{a}\partial_\mu^2 \psi (x),
\]
\vspace{-7mm}
\begin{equation} \label{12}
\end{equation}
\vspace{-7mm}
\[
\Psi_\mu (x)=\psi_\mu \equiv -\frac{b}{a}\partial_\mu \psi (x) .
\]
The wave function $\Psi (x)$ represents, in Eq. (12), the direct sum
of bispinor $\psi (x)$, bispinor $\widetilde{\psi} (x)$,
vector-bispinor $\psi _\mu (x)$, and transforms under the Lorentz
group as
\[
[(1/2,0)\oplus(0,1/2)]\oplus[(1/2,0)\oplus(0,1/2)]\oplus\{(1/2,1/2)\otimes
[(1/2,0)\oplus(0,1/2)]\}.
\]
To obtain the first order relativistic wave equation from Eq. (9),
we introduce the elements of the entire algebra $\varepsilon ^{A,B}$
(see, for example, \cite{Kruglov1}) obeying the multiplication rule
and having matrix elements as follows:
\begin{equation}
\varepsilon ^{M,A}\varepsilon ^{B,N}=\delta _{AB}\varepsilon
^{M,N},\hspace{0.5in} \left( \varepsilon ^{M,N}\right) _{AB}=\delta
_{MA}\delta _{NB},  \label{13}
\end{equation}
$A,B,M,N=0,\widetilde{0},\mu$ ($\mu=1,2,3,4$). The matrix
$\varepsilon ^{M,N}$ consists of zeros and only one element is
unity, where row $M$ and column $N$ cross.

Using Eq. (11)-(13), Eq. (9) takes the form of the first order
equation
\[
\partial _\mu \left(\varepsilon ^{\mu,0 }+ \varepsilon ^{\widetilde{0},\mu}
- \varepsilon ^{0,\mu}+ \varepsilon^{0,0}\gamma_\mu +
\varepsilon^{0,\widetilde{0}}\gamma_\mu \right)_{AB}\Psi _B(x)
\]
\vspace{-7mm}
\begin{equation} \label{14}
\end{equation}
\vspace{-7mm}
\[
+ \left( \frac{a}{b}\varepsilon ^{\mu ,\mu }+
\frac{c}{a}\varepsilon ^{0,0} + b\varepsilon
^{\widetilde{0},\widetilde{0}}\right) _{AB}\Psi _B(x)=0 .
\]
It is implied that $\gamma$-matrices act on the bispinor indexes.
There is a summation over repeated indices in subscripts and
superscripts of the matrices. Now we introduce the matrices
\begin{equation}
\Gamma _\mu =\left(\varepsilon ^{\mu,0 }+ \varepsilon
^{\widetilde{0},\mu}- \varepsilon ^{0,\mu}\right)\otimes I_4 +
\left( \varepsilon^{0,0}+
\varepsilon^{0,\widetilde{0}}\right)\otimes\gamma_\mu , \label{15}
\end{equation}
\begin{equation}
 M =\frac{c}{a}P_0 +
 b\widetilde{P}_0 + \frac{a}{b}P_1 , \label{16}
\end{equation}
\begin{equation}
P_0=\varepsilon ^{0,0}\otimes I_4 ,~~~~\widetilde{P}_0=\varepsilon
^{\widetilde{0},\widetilde{0}}\otimes I_4 ,~~~~P_1=\varepsilon ^{\mu
,\mu }\otimes I_4 . \label{17}
\end{equation}
Unit four-dimensional matrix $I_4$ acts on bispinor indexes and
matrices $\varepsilon ^{M ,N }$ ($M, N=0,\widetilde{0},\mu$) act on
scalar and vector indexes. It should be mentioned that matrices
$P_0$, $\widetilde{P}_0$, $P_1$ are projection matrices obeying the
relation $P^2=P$. The $M$ is the diagonal and Hermitian matrix, $M =
M^+$, but the matrix $\Gamma _\mu$ is not Hermitian matrix.

Taking into account Eq. (11)-(13), Eq. (14) becomes
\begin{equation}
\left( \Gamma _\mu \partial _\mu +M\right) \Psi (x)=0 . \label{18}
\end{equation}
Eq. (18) obtained, is the relativistic 24-component wave equation of
the first order and is convenient for different applications.

Now, we derive the mass spectrum of Eq. (18). In the frame of
references where a particle is at the rest, Eq. (18), in the
momentum space, becomes
\begin{equation}
\left( p_0\Gamma _4 - M\right) \Psi (p_0)=0 , \label{19}
\end{equation}
where $p_0$ is the energy of a particle. Here we use one particle
(quantum mechanical) interpretation of Eq. (18) and do not discuss
the second quantized theory. Masses of particles (see
\cite{Gel'fand}) are given by the relation $m_i= \lambda_i^{-1}$,
where $\lambda_i$ are eigenvalues of the matrix $M^{-1}\Gamma_4$ or
$m_i=\overline{\lambda}_i$, where $\overline{\lambda}_i$ are
eigenvalues of the matrix $M\Gamma_4^{-1}$. From Eq. (15), (16), one
may find expressions
\begin{equation}
 M^{-1} =\frac{a}{c}P_0 +
 \frac{1}{b}\widetilde{P}_0 + \frac{b}{a}P_1 , \label{20}
\end{equation}
\begin{equation}
K\equiv M^{-1}\Gamma_4=\left(\frac{b}{a}\varepsilon ^{4,0 }+
\frac{1}{b}\varepsilon ^{\widetilde{0},4}- \frac{a}{c}\varepsilon
^{0,4}\right)\otimes I_4 + \left(\frac{a}{c}\varepsilon^{0,0}+
\frac{a}{c}\varepsilon^{0,\widetilde{0}}\right)\otimes\gamma_4 ,
\label{21}
\end{equation}

With the help of Eq. (13), it is easy to verify that the matrix $K$
obeys the equation as follows:
\begin{equation}
 cK^3I_6\otimes\gamma_4 -aK^2 +bKI_6\otimes\gamma_4-\Lambda=0 , \label{22}
\end{equation}
where
\begin{equation}
I_6\otimes\gamma_4 =\left(\varepsilon
^{0,0}+\varepsilon^{\widetilde{0},\widetilde{0}}+\varepsilon
^{\mu,\mu}\right)\otimes\gamma_4 , \label{23}
\end{equation}
\begin{equation}
\Lambda =P_0 +\widetilde{P}_0+\varepsilon ^{4,4}\otimes I_4 .
\label{24}
\end{equation}
The matrix $\Lambda$ is the projection matrix, so that
$\Lambda^2=\Lambda$. It follows from Eq. (22), that as the matrix
$I_6\otimes\gamma_4$ possesses eigenvalues $\pm 1$, the matrix
$\Lambda$ possesses eigenvalues $1$, eigenvalues of the matrix $K$,
$\lambda=\{\lambda_i\}$, obey the equation
\begin{equation}
 \pm c\lambda^3 -a\lambda^2 \pm b\lambda-1=0 . \label{25}
\end{equation}
As a result, taking into consideration the relation
$m_i=1/\lambda_i$, the mass spectrum of Eq. (18) is given by
solutions of the equation
\begin{equation}
 m_i^3 \mp b m_i^2 +a m_i \mp c =0 . \label{26}
\end{equation}
Positive solutions to Eq. (26), $m_i$, are connected with the
coefficients $a$, $b$,and $c$, possessing the flavor $S_3$ symmetry,
by relationships (10). So, the mass spectrum of Eq. (18) and Eq. (9)
is the same.

As the diagonal mass matrix $M$ is not proportional to unit matrix,
it is convenient for some applications to get the equivalent form of
Eq. (18). Multiplying Eq. (18) by the non-singular matrix
\[
N=P_0 + \frac{c}{ab}{P}_0 + \frac{cb}{a^2}P_1 ,
\]
and introducing the new matrix
\begin{equation}
\overline{\Gamma}_\mu=N\Gamma_\mu=\left(\frac{cb}{a^2}\varepsilon
^{\mu,0 }+ \frac{c}{ab}\varepsilon ^{\widetilde{0},\mu}- \varepsilon
^{0,\mu}\right)\otimes I_4 + \left( \varepsilon^{0,0}+
\varepsilon^{0,\widetilde{0}}\right)\otimes\gamma_\mu , \label{27}
\end{equation}
Eq. (18) transforms to the standard form of the relativistic wave
equation
\begin{equation}
\left( \overline{\Gamma}_\mu \partial _\mu +m\right) \Psi (x)=0 .
\label{28}
\end{equation}
We took into consideration the equalities
\[
NM=\frac{c}{a}I_{24},~~m\equiv \frac{c}{a}=\frac{m_1 m_2 m_3}{m_1
m_2 + m_1 m_3 + m_2 m_3},
\]
where $I_{24}$ is the unit $24\times 24$-matrix, and $m$ is a
reduced mass. Notice that the wave function $\Psi (x)$ remains the
same (Eq. (11)) in Eq. (18),(28).

\section{Mass and Spin Projection Operators}

In the momentum space Eq. (28) becomes
\begin{equation}
\left( i\breve{p}+m\right) \Psi(p)=0 , \label{29}
\end{equation}
where $\breve{p}=p_\mu\overline{\Gamma}$. One may verify, with the
help of Eq. (A6)-(A7) (see Appendix A), that the $\breve{p}$ obeys
the matrix equation as follows:
\begin{equation}
\breve{p}\left(\breve{p}^2+m^2 \right) \biggl[c^2\breve{p}^4
-m^2\breve{p}^2\left(p^2 d+ c^2\right) -m^4p^6 \biggr] =0 ,
\label{30 }
\end{equation}
where $d=m_1^2m_2^2+ m_1^2m_3^2 + m_3^2m_2^2$. From Eq. (30), we
find the projection operator
\begin{equation}
\Pi_+=\frac{\breve{p}\left(\breve{p}+im \right) \left[c^2\breve{p}^4
-m^2\breve{p}^2\left(p^2d+ c^2\right) -m^4p^6
\right]}{2m^6\left(p^6- p^2d -2c^2\right)} . \label{31 }
\end{equation}
It is not difficult to verify, with  the help of Eq. (30)  that the
the projection operator (31) obeys the equation $\Pi_+^2=\Pi_+$. To
find the projection operator corresponding to negative energy,
$\Pi_-$, one has to make the replacement $p\rightarrow -p$ in Eq.
(31). Every column of the matrix $\Pi_+$ is a solution to Eq. (29),
and after acting on $24$-dimensional vector $\chi$, one finds the
solution to Eq. (29), $\Psi(p)=\Pi_+\chi$. The matrix (31) also
represents the density matrix for the impure spin state.

To obtain the spin projection operators, we consider the generators
of the Lorentz group in the 24-dimensional representation which are
given by
\begin{equation}
J_{\mu \nu }=\left(
\begin{array}{ccc}
0 & 0 & 0 \\
0 & 0 & 0 \\
0 & 0 &  J_{\mu \nu }^{(1)}
\end{array}
\right)\otimes I_4+I_6 \otimes J_{\mu\nu}^{(1/2)} ,
 \label{32}
\end{equation}
where \cite{Kruglov1}
\begin{equation}
J_{\mu\nu}^{(1)}= \varepsilon^{\mu,\nu}-\varepsilon^{\nu,\mu} ,
 \label{33}
\end{equation}
\begin{equation}
J_{\mu\nu}^{(1/2)}= \frac{1}{4}\left( \gamma_\mu
\gamma_\nu-\gamma_\nu \gamma_\mu \right) .
 \label{34}
\end{equation}
Unit 6-dimensional matrix $I_6$ is defined in the space
$(0,\widetilde{0},\mu$), and the generators of the Lorentz group in
four-dimensional vector and bispinor spaces are $J_{\mu\nu}^{(1)}$,
$J_{\mu\nu}^{(1/2)}$, respectively. It is easy to verify that
generators (32)-(34) satisfy the commutation relations
\begin{equation}
\left[ J_{\mu \nu },J_{\alpha \beta}\right] =\delta _{\nu \alpha
}J_{\mu \beta}+\delta _{\mu \beta }J_{\nu \alpha}-\delta _{\nu \beta
}J_{\mu \alpha}-\delta _{\mu \alpha }J_{\nu \beta} . \label{35}
\end{equation}
In our euclidian metric, the antisymmetric parameters of the Lorentz
group $\omega_{mn}$ ($m,n=1,2,3$) are real, and $\omega_{m4}$ are
imaginary. The relativistic form-invariance of Eq. (28) follows from
the commutation relation \cite{Gel'fand}
\begin{equation}
\left[ \overline{\Gamma} _\lambda ,J_{\mu \nu }\right] =\delta
_{\lambda \mu }\overline{\Gamma} _\nu -\delta _{\lambda \nu
}\overline{\Gamma} _\mu , \label{36}
\end{equation}
which is valid for matrices (27), (32).

The operator of the spin projection on the direction of the
momentum $\textbf{p}$ is given by (see, for instance,
\cite{Kruglov1})
\begin{equation}
\sigma_p=-\frac{i}{2|\textbf{p}|}\epsilon_{abc}\textbf{p}_a J_{bc} ,
\label{37}
\end{equation}
where $|\textbf{p}| =\sqrt{p_1^2 +p_2^2+p_3^2}$, $\epsilon_{abc}$
($a,b,c=1,2,3$) is an antisymmetric tensor Levy-Civita. With the
help of Eq. (36), one may verify that the operator of Eq. (28) in
the momentum space $(i\overline{\Gamma}_\mu p_\mu +m)$ commutes with
the spin operator (37): $[i\overline{\Gamma}_\mu p_\mu
+m,\sigma_p]=0$. This guaranties the existence of the common
solutions of the Eq. (29) in the momentum space, $\Psi_{s_p} (p)$,
and the equation
\begin{equation}
\sigma_p \Psi_{s_p}(p)=s_p \Psi_{s_p}(p) , \label{38}
\end{equation}
where $s_p=\pm 1/2$. With the aid of Eq. (13), (B5) (see Appendix
B), one can verify that the matrix equation
\begin{equation}
\left(\sigma_p^2-\frac{1}{4}\right)\left(\sigma_p^2-\frac{9}{4}\right)=0
\label{39}
\end{equation}
is valid. This equation allows us to construct two projection
operators corresponding to spin $1/2$ and $3/2$. Indeed, the
representation of the Lorentz group corresponding to the
$24$-dimensional wave function (11), contains also the spin-3/2
component. It is important to show that the wave equation (28) (or
(18)) describes only pure spin-1/2 states. For this purpose, we
consider both projection operators corresponding to spin $1/2$ and
$3/2$. Exploring the method \cite{Fedorov}, one obtains the
projection operators
\begin{equation}
P_{\pm1/2}=\mp \frac{1}{2}\left(\sigma_p\pm \frac{1}{2}
\right)\left(\sigma_p^2-\frac{9}{4}\right),~~ P_{\pm 3/2}=\mp
\frac{1}{6}\left(\sigma_p\pm \frac{3}{2}
\right)\left(\sigma_p^2-\frac{1}{4}\right)  \label{40}
\end{equation}
extracting spin projections $\pm 1/2$ and  $\pm 3/2$. One can verify
that $P_{s_p}^2=P_{s_p}$ ($s_p=\pm 1/2, \pm 3/2$). With the help of
Eq. (B6), we find the relations
\begin{equation}
P_{\pm 3/2}\breve{p}=\breve{p}P_{\pm 3/2}=0,~~~~P_{\pm 3/2}\Pi_\pm
=\Pi_\pm P_{\pm 3/2}=0 , \label{41}
\end{equation}
but $P_{\pm 1/2}\Pi_\pm \neq 0$. Therefore, the wave equation (28)
(or (18)) describes only pure spin-1/2 states. Eq. (39), (40) lead
to the equation
\begin{equation}
\sigma_p P_{s_p}=s_p P_{s_p} . \label{42}
\end{equation}
Eq. (42) shows that the wave function $\Psi_{s_p} (p)=P_{s_p}\chi $,
($\chi$ being an arbitrary nonzero $24$-dimensional column) is the
solution of Eq. (38). The projection operator corresponding to
common solutions to Eq. (29), (38) is given by
\begin{equation}
\rho_{\pm 1/2} (p)=\Pi_{+} P_{\pm 1/2} . \label{43}
\end{equation}
The projection operator (43) extracts states with positive energy of
particles and spin projections $\pm 1/2$. The $\rho_{\pm 1/2}$ is
also the density matrix for pure spin states ($s_p=\pm 1/2$).

\section{Non-Minimal Electromagnetic Interactions of Fermions}

Now, we obtain the density of the electromagnetic current.  For this
purpose, Eq. (2) is rewritten as follows:
\begin{equation}
\partial_\mu \partial_\nu^2\overline{\psi }(x)\gamma_\mu
+a\partial_\mu \overline{\psi }(x)\gamma_\mu
-b\partial_\nu^2\overline{\psi }(x)-c \overline{\psi}(x)=0.
\label{44}
\end{equation}
Following the well-known procedure, multiplying Eq. (9) (from left)
by $\overline{\psi }(x)$, and Eq. (44) (from right) by $\psi (x)$,
and adding them, one finds
\[
\overline{\psi }(x)\gamma_\mu \partial_\mu \partial_\nu^2 \psi (x)+
\left(\partial_\mu \partial_\nu^2 \overline{\psi
}(x)\right)\gamma_\mu\psi (x)
\]
\begin{equation}
+a\left[ \overline{\psi }(x)\gamma_\mu \partial_\mu \psi (x) +
\left( \partial_\mu \overline{\psi }(x)\right)\gamma_\mu \psi
(x)\right] \label{45}
\end{equation}
\[
+ b\left[ \overline{\psi }(x)\partial_\nu^2 \psi (x)
-\left(\partial_\nu^2 \overline{\psi }(x) \right)\psi (x)\right] = 0
.
\]
Identifying the left side of Eq. (45) with $(-ia)\partial_\mu j_\mu
(x)$, we obtain the electric current density
\[
j_\mu (x)=i\overline{\psi }(x)\gamma_\mu \psi(x)+
i\frac{b}{a}\left[\overline{\psi }(x)\partial_\mu \psi(x)
-\left(\partial_\mu\overline{\psi}(x)\right) \psi(x)\right]
\]
\begin{equation}
+\frac{i}{a}[\overline{\psi }(x)\gamma_\mu
\partial_\nu^2\psi(x)
+ \left(\partial_\nu^2\overline{\psi }(x)\right)\gamma_\mu \psi(x)
-\left(\partial_\nu\overline{\psi}(x)\right)\gamma_\nu
\partial_\mu \psi(x)
\label{46}
\end{equation}
\[
- \left(\partial_\mu\overline{\psi}(x)\right)\gamma_\nu
\partial_\nu \psi(x)+ \left(\partial_\nu\overline{\psi}(x)\right)\gamma_\mu
\partial_\nu \psi(x) ] ,
\]
so that the conservation of the four-vector current density
$\partial_\mu j_\mu (x)=0$ holds. The current $j_\mu (x)$ consists
of the usual Dirac current $j^D_\mu (x)=i\overline{\psi
}(x)\gamma_\mu \psi(x)$ and additional convective terms.

Substituting the derivatives in Eq. (18), $\partial _\mu \rightarrow
D_\mu =\partial _\mu -ieA_\mu $ ($A_\mu $ is the four-vector
potential of the electromagnetic field), one can obtain the minimal
interaction with an electromagnetic field. Now we introduce the
non-minimal electromagnetic interaction by considering the matrix
equation
\begin{equation}
\biggl [\Gamma _\mu D_\mu +\frac i2 \mathcal{F}_{\mu \nu}\Gamma
_{\mu \nu}\left( \kappa _0P_0 +\widetilde{\kappa} _0\widetilde{P}_0
+\kappa _1 P_1\right) +M\biggr ]\Psi (x)=0 ,\label{47}
\end{equation}
where $\mathcal{F}_{\mu \nu }=\partial _\mu A_\nu -\partial _\nu
A_\mu $ is the strength of the electromagnetic field, and
$\kappa_0$, $\widetilde{\kappa}_0$, $\kappa_1$ are parameters
which characterize anomalous electromagnetic interactions of
fermions, and
\[
\Gamma _{\mu \nu }=\Gamma _\mu \Gamma _\nu -\Gamma _\nu \Gamma
_\mu
\]
\begin{equation}
=\left(\varepsilon^{0,\widetilde{0}}+\varepsilon^{0,0}\right)
\otimes\left(\gamma_\mu\gamma_\nu-\gamma_\nu\gamma_\mu\right)
+\left[\varepsilon^{\mu,0}+\varepsilon^{\mu,\widetilde{0}}
\right]\otimes\gamma_\nu \label{48}
\end{equation}
\[
-\left[\varepsilon^{\nu,0}+\varepsilon^{\nu,\widetilde{0}}
\right]\otimes\gamma_\mu
+\left(\varepsilon^{\nu,\mu}-\varepsilon^{\mu,\nu}\right)\otimes
I_4.
\]
The projection operators $P_0$, $\widetilde{P}_0$, $P_1$  obey the
relations: $P_0 \widetilde{P}_0=P_0 P_1= \widetilde{P}_0 P_1=0$,
$P_0+\widetilde{P}_0+P_1=I_{24}$. Eq. (47) is the relativistic wave
equation for interacting fermions which is form-invariant under the
Lorentz transformations. The tensor form of Eq. (47) follows from
Eq. (11), (15)-(17), (48):
\[
 \gamma_\mu D_\mu\left[\psi (x)+\widetilde{\psi} (x)\right]
 +i\mathcal{F}_{\mu \nu} \gamma_\mu \gamma_\nu
\left[\kappa_0 \psi (x)+\widetilde{\kappa}_0 \widetilde{\psi}
(x)\right]
\]
\vspace{-7mm}
\begin{equation} \label{49}
\end{equation}
\vspace{-7mm}
\[
-D_\mu\psi_\mu (x)+\frac{c}{a}\psi (x)=0 ,
\]
\begin{equation}
D _\mu \psi (x) +i\mathcal{F}_{\mu \nu}\gamma_\nu
\left[\kappa_0\psi (x)+\widetilde{\kappa}_0 \widetilde{\psi} (x)
\right] + \left(\frac{a}{b}\delta_{\mu\nu}-i\kappa_1
\mathcal{F}_{\mu \nu}\right)\psi_\nu (x)=0 ,
 \label{50}
\end{equation}
\begin{equation}
D _\mu \psi_\mu (x) + b\widetilde{\psi} (x)=0 .
 \label{51}
\end{equation}
Eq. (49)-(51) can be used for the phenomenological applications in
the case of anomalous electromagnetic interactions of fermions. The
system of Eq. (49)-(51) defines bispinors $\psi (x)$,
$\widetilde{\psi} (x)$, and vector-bispinor $\psi_\nu (x)$. Eq. (50)
may be considered as a matrix equation for vector-bispinor $\psi_\nu
(x)$. Finding $\widetilde{\psi} (x)$, and vector-bispinor $\psi_\nu
(x)$ from Eq. (51), (50) and substituting them into Eq. (49), one
may obtain an equation for bispinor $\psi (x)$. This equation
contains the interaction with the anomalous magnetic moments of
particles. To figure out the physical interpretation of introduced
constants $\kappa_0$, $\kappa_1$, $\widetilde{\kappa}_0$, one needs
to consider the non-relativistic limit of the theory. We leave this
for further investigations.

Let us consider the Hamiltonian form of Eq. (47). This form is
convenient for determination of dynamical variables of the wave
function $\Psi (x)$ and for second quantization of fields under
consideration. Eq. (47) can be rewritten as
\begin{equation}
i\Gamma _4 \partial_t \Psi (x)=\biggl [\Gamma_m D_m +eA_0\Gamma_4
+\frac i2 \mathcal{F}_{\mu \nu }\Gamma _{\mu \nu }\left( \kappa
_0P_0 +\widetilde{\kappa} _0\widetilde{P}_0 +\kappa _1 P_1\right)
+M\biggr ]\Psi (x) .\label{52}
\end{equation}
With the aid of Eq. (13), one may verify that the matrix $\Gamma_4$
obeys the equation as follows:
\begin{equation}
\Gamma_4^4=\Lambda ,
 \label{53}
\end{equation}
where the projector operator $\Lambda$ is given by Eq. (24). To
separate the canonical and non-canonical parts of the wave function,
we define the projection operator
\begin{equation}
\Pi=1-\Lambda ,
 \label{54}
\end{equation}
where $1\equiv I_{24}$ is the unit $24\times 24$-matrix. Projection
operators $\Lambda$, $\Pi$ obey equations
\begin{equation}
\Pi +\Lambda=1 ,~~~~\Pi \Lambda =\Lambda\Pi =0
,~~~~\Lambda^2=\Lambda ,~~~~\Pi^2=\Pi.
 \label{55}
\end{equation}
Introducing the auxiliary wave functions
\begin{equation}
\phi(x)=\Lambda\Psi(x),~~~~\chi (x)=\Pi \Psi (x) ,
 \label{56}
\end{equation}
so that $\phi(x)+\chi (x)=\Psi (x)$, multiplying Eq. (52) by
$\Gamma_4^3$, and taking into account Eq. (53), one obtains
\begin{equation}
i \partial_t \phi (x)=eA_0\phi (x)+\Gamma_4^3\biggl (\Gamma_m D_m
+\sigma +M\biggr )\left[ \phi (x)+\chi (x)\right] ,\label{57}
\end{equation}
where we have introduced the operator
\begin{equation}
\sigma=\frac i2 \mathcal{F}_{\mu \nu }\Gamma _{\mu \nu }\left(
\kappa _0P_0 +\widetilde{\kappa} _0\widetilde{P}_0 +\kappa _1
P_1\right) . \label{58}
\end{equation}
Acting on Eq. (52) with the operator $\Pi$, and using the equations
\[
\Pi \Gamma_4=0 ,~~~~\Pi M=\frac{a}{b}\Pi ,~~~~\Pi \Gamma_m \Pi =0
,
\]
we obtain the equation for auxiliary wave function $\chi (x)$
\begin{equation}
\frac{a}{b}\chi (x) + \Pi \Gamma_m D_m \phi(x) +\Pi\sigma \left([
\phi (x)+\chi (x)\right]=0 . \label{59}
\end{equation}
Expressing the auxiliary wave function $\chi (x)$ from Eq. (59), and
replacing it into Eq. (57), one obtains the equation in the
Hamiltonian form:
\[
i \partial_t \phi (x)={\cal H}\phi (x) ,
\]
\vspace{-7mm}
\begin{equation} \label{60}
\end{equation}
\vspace{-7mm}
\[
 {\cal H}=eA_0 +\Gamma_4^3\biggl (\Gamma_m D_m
+\sigma +M\biggr )\left[1-\left(\frac{a}{b}+\Pi\sigma \right)^{-1}
\Pi \left(\Gamma_m D_m +\sigma\right) \right] .
\]
The wave function $\phi (x)$ entering Eq. (60), possesses $12$
components, and the auxiliary wave function $\chi (x)$ also has $12$
components. Indeed, for the relativistic description of spin-1/2
fields (four components for each field) with three mass states, $12$
components are needed, and, therefore, the dynamical wave function
$\phi (x)$, possesses $12$ components.

At the particular case, for point-like particles, without anomalous
interactions, $\sigma=0$, one finds the Hamiltonian from Eq. (60) as
follows:
\begin{equation}
 {\cal H}=eA_0 +\Gamma_4^3\biggl (\Gamma_m D_m
+M\biggr )\left(1- \frac{b}{a} \Pi \Gamma_m D_m \right) . \label{61}
\end{equation}

\section{Conclusion}

We have considered the generalized Dirac equation of the third
order, describing fermions possessing the permutation symmetry
$S_3$. The 24-dimensional matrix form of the first order was
obtained which is convenient for different applications. Although
the dimension of the matrices is high ($24\times 24$), we have found
minimal equations for matrices. This allowed us to obtain the
projection operators extracting states with definite energy and spin
projections. The density of the electromagnetic current was
obtained, and we have introduced three phenomenological parameters
characterizing anomalous non-minimal electromagnetic interactions of
fermions. It is of interest to clear up the physical interpretation
of constants introduced. This and other questions we leave for
further investigations.

Dynamical and non-dynamical variables were separated and
quantum-mechanical Hamiltonian was derived. Fist order matrix wave
equations obtained are convenient for second quantization and for
calculations of different electrodynamic processes.

Non-trivial differences between this three mass case and the
previously studied two mass generalized Dirac equation
\cite{Kruglov} is in the dimension of matrices of wave equation and
corresponding minimal equations. We mention that in this three mass
case it is not trivial to obtain the Hermitianizing matrix $\eta$ in
our $24$- representation case and corresponding Lagrangian.

The HD field theory considered includes the ghost state which
requires to introduce indefinite metrics. It should be noted that
this negative attribute appears in many HD field theories.

One of important unanswered questions is the mechanism of mass
generations of fermions (see \cite{Koide} and references there).
Here, we have introduced three independent parameters which specify
the masses and do not explain their appearance. It seems interesting
to follow the Barut \cite{Barut1} idea about the electromagnetic
nature of fermion masses. This, however, requires the further
investigations of the approach considered.

\vspace{5mm}
\begin{center}
\textbf{Appendix A}
\end{center}
\vspace{5mm}

For a convenience, we represent the matrix $\breve{p}$ of Eq. (29)
as follows
\[
\breve{p}=I_{(0)}\otimes\widehat{p}+I_{(1)}\otimes I_4 ,
\hspace{2.0in}\left( A1\right)
\]
where
\[
I_{(0)}= \varepsilon^{0,0}+ \varepsilon^{0,\widetilde{0}},
~~I_{(1)}=p_\mu\left(\alpha\varepsilon ^{\mu,0 }+ \beta\varepsilon
^{\widetilde{0},\mu}- \varepsilon ^{0,\mu}\right),
\hspace{1.0in}\left( A2\right)
\]
\[
\alpha=\frac{cb}{a^2},~~~~\beta=\frac{c}{ab},~~~~\widehat{p}=p_\mu\gamma_\mu.
\hspace{2.0in}\left( A3\right)
\]
Using Eq. (13) one may prove simple relationships:
\[
I_{(0)}I_{(1)}I_{(0)}=0,~~~~I_{(1)}^3=-\alpha
p^2I_{(1)},~~~~I_{(1)}I_{(0)}I_{(1)}=\alpha\left(\beta
-1\right)I_{(2)},
\]
\[
I_{(2)}=p_\mu p_\nu \varepsilon^{\mu,\nu},~~~~ I_{(2)}^2=p^2I_{(2)},
\]
\[
I_{(2)}I_{(0)}=I_{(0)}I_{(2)}=0,~~~~I_{(1)}I_{(2)}I_{(1)}=p^2\left(
I_{(1)}^2+\alpha I_{(2)}\right),\hspace{1.0in}\left( A4\right)
\]
\[
I_{(2)}I_{(1)}^2=-\alpha p^2I_{(2)},
~~~~I_{(0)}I_{(1)}^2I_{(0)}=\alpha(\beta-1)p^2I_{(0)},
\]
\[
I_{(1)}^2I_{(0)}I_{(1)}+I_{(1)}I_{(0)}I_{(1)}^2=\alpha(\beta-1)p^2I_{(1)}.
\]
One may check, with the help of Eq. (13),(A4), that the minimal
equation
\[
\breve{p}^7 +(2\alpha-1)p^2\breve{p}^5+\alpha(\alpha-2\beta)
p^4\breve{p}^3 -\alpha^2\beta^2p^6\breve{p}=0 . \hspace{1.0in}\left(
A5\right)
\]
is valid. It should be noted that the dispersion equation (4) can be
rewritten as
\[
m^6 -(2\alpha-1)m^4p^2+\alpha(\alpha-2\beta)m^2 p^4
+\alpha^2\beta^2p^6=0 . \hspace{1.0in}\left( A6\right)
\]
With the help of Eq. (A6), Eq. (A5) may be represented as follows:
\[
\breve{p}\left(\breve{p}^2-\frac{m^2}{m_1^2}p^2\right)\left(\breve{p}^2-\frac{m^2}
{m_2^2}p^2\right) \left(\breve{p}^2-\frac{m^2}{m_3^2}p^2\right) =0 .
\hspace{1.0in}\left( A7\right)
\]
Although the $\breve{p}$ is $24\times 24$-matrix it obeys the simple
minimal equation.

\vspace{5mm}
\begin{center}
\textbf{Appendix B}
\end{center}
\vspace{5mm}

The spin operator (37) can be written as
\[
\sigma_p=\sigma_p^{(1)}\otimes I_4+ I_6\otimes\sigma_p^{(1/2)},
\hspace{1.0in}\left( B1\right)
\]
where
\[
\sigma_p^{(1)}=-\frac{i}{2|\textbf{p}|}\epsilon_{abc}\textbf{p}_a
J_{bc}^{(1)} ,~~~~\sigma_p^{(1/2)}=
-\frac{i}{2|\textbf{p}|}\epsilon_{abc}\textbf{p}_a J_{bc}^{(1/2)}.
\hspace{1.0in}\left( B2\right)
\]
The operators $\sigma_p^{(1)}$, $\sigma_p^{(1/2)}$ correspond to
spin-1 and spin-1/2 representations of the Lorentz group. Using Eq.
(33), (34), we obtain
\[
\left(\sigma_p^{(1)}\right)^2=\varepsilon^{m,m}-\frac{p_a
p_b}{\textbf{p}^2}\varepsilon^{a,b},
~~\left(\sigma_p^{(1/2)}\right)^2=\frac{1}{4}.\hspace{1.0in}\left(
B3\right)
\]
so, that the spin operators obey the simple matrix equations
\[
\sigma_p^{(1)}\left(\sigma_p^{(1)}-1\right)\left(\sigma_p^{(1)}+1\right)=0,~~~~
\left(\sigma_p^{(1/2)}-\frac{1}{2}\right)\left(\sigma_p^{(1/2)}+\frac{1}{2}\right)=0.
\hspace{1.0in}\left( B4\right)
\]
From Eq. (B1), (B3), one finds
\[
\left(\sigma_p\right)^2=\frac{1}{4}+\left(\sigma_p^{(1)}\right)^2\otimes
I_4+ 2\sigma_p^{(1)}\otimes\sigma_p^{(1/2)}. \hspace{1.0in}\left(
B5\right)
\]
Using Eq. (B3), (B5), we obtain the useful relation
\[
\left[\left(\sigma_p\right)^2-\frac{1}{4}\right]\breve{p}=0.
\hspace{1.0in}\left( B6\right)
\]

\end{document}